\documentclass[12pt, a4paper]{article}
\usepackage{amsfonts}
\usepackage{amsmath}
\usepackage{amssymb}
\usepackage{amsthm}
\usepackage{rotating}
\usepackage{color}
\usepackage[table]{xcolor}
\usepackage{endnotes}
\usepackage{harvard}
\usepackage{booktabs}
\usepackage{threeparttable}
\usepackage{longtable}
\usepackage{float}

\newcommand{\blind}{0}

\begin{document}

\def\spacingset#1{\renewcommand{\baselinestretch}%
{#1}\small\normalsize} \spacingset{1}


\if0\blind
{

\title{Measuring educational attainment as a continuous variable: a new database (1970-2010)}
\author{Vanesa Jorda\thanks{The research leading to these results has received funding from the Ministerio de Econom\'ia y Competitividad (project ECO2013-48326-C2-2-P). The authors are also grateful to Markus Jantti, Koen Decancq, Rolf Aaverge, Stefan Sperlich, Jose Maria Sarabia and participants of the X Winter School on Inequality and Social Welfare Theory, the Sixth ECINEQ meeting and the CSB Seminar at the University of Antwerp for their valuable comments.} \hspace{.2cm}\\
    UNU-WIDER/ Department of Economics, University of Cantabria\\
    and \\
    Jose M. Alonso \\
    Department of Economics, University of Cantabria}
  \maketitle
}

\fi

\if1\blind
{
  \bigskip
  \bigskip
  \bigskip
  \begin{center}
    {\LARGE Measuring educational attainment as a continuous variable: a new database (1970-2010)}
\end{center}
  \medskip
} \fi

\bigskip

\begin{abstract}

In this paper, we introduce a new comprehensive data set on educational attainment and inequality measures of education for 142 countries over the period 1970 to 2010. Most of the previous attempts to measure educational attainment have treated education as a categorical variable, whose mean is computed as a weighted average of the official duration of each cycle and attainment rates, thus omitting differences in educational achievement within levels of education. This aggregation into different groups may result in a loss of information introducing, therefore, a potential source of measurement error. We explore here a more nuanced alternative to estimate educational attainment, which considers the continuous nature of the educational variable. This `continuous approach' allows us to impose more plausible assumptions about the distribution of years of schooling within each level of education, and to take into account the right censoring of the data in the estimation, thus leading to more accurate estimates of educational attainment and education inequality. These improved series may help to better understand the role of education on different socio-economic aspects, such as quality of life and human capital formation.

\end{abstract}


\vfill

\newpage
\spacingset{1.45} 
\section{Introduction}

The role of education on quality of life and human capital formation is an enduring issue in development economics. Although income variables were traditionally considered the main indicators to measure well-being levels, characterizing therefore quality of life as a purely economic process, many scholars have emphasized that the assessment of well-being should include other non-economic dimensions that may be equally relevant (Sen, \citeyear*{sen88}; Stiglitz et al., \citeyear*{Stiglitz10}). Among those non-economic dimensions, education is acknowledged to be a key factor to measure well-being (Acemoglu and Angrist, \citeyear*{acemoglu01}; Oreopoulos and Salvanes, \citeyear*{Oreopoulos11}). At the same time, education is regarded as an important means to build up human capital, thus being a key factor for economic growth and development (Lucas, \citeyear*{LUCAS19883}; Becker et al., \citeyear*{Becker1990}). Due to this key role of education in social and economic progress, it is not surprising  that development economists (among other scholars) have devoted considerable efforts towards providing consistent estimates of educational attainment for a large number of countries over reasonably long periods of time (see, for example, Cohen and Soto, \citeyear*{cohen07}; Morrisson and Murtin, \citeyear*{morrisson09}; Barro and Lee, \citeyear*{barro2013}).

Most of the previous attempts to measure educational attainment have focused on estimating the mean years of schooling (henceforth, MYS), since years of schooling have been often considered a good proxy for the underlying phenomenon of gaining knowledge.  One of the major problems when estimating MYS is the lack of microdata on the exact number of years of schooling completed by an individual. Hence, individuals are often grouped into four broad educational levels according to their educational attainment, namely no schooling, primary, secondary and tertiary education, which are then divided into complete and incomplete, whether the educational cycle has been finished or not. Due to this particular structure of the data, education is treated as a categorical variable whose mean is computed as a weighted average of the official duration of each cycle and attainment rates. This approach implicitly \emph{discretizes} the variable \emph{time that individuals attend school until leaving the educational system} to obtain an easily measurable indicator of educational outcomes. However, educational attainment is not a discrete variable but a continuous one, and the aggregation into different groups may result in a loss of information, thus introducing a potential source of measurement error.

Measurement error associated with treating education as a discrete variable may arise from different sources. First, completion rates used to break down educational levels into complete and incomplete are - in most cases - estimated. Next, the exact years of schooling are known only for those individuals classified into complete levels, corresponding to the official duration of each cycle. However, there is no information about the distribution of education within incomplete educational stages. To deal with this data limitation, the same arbitrary number of years of schooling is given to all individuals who did not complete a particular educational level.\footnote{For instance, MYS provided by UNESCO are calculated assuming that people who have not finished the educational cycle have successfully completed half of the official duration of such a level. Then, these estimates would be always biased except in the hypothetical case that population distributes uniformly across years within each level of education, which would be a coincidence rather than the norm.} This arbitrary choice may introduce some measurement error and, more importantly, it may be difficult to determine the direction of potential biases. As a consequence, the assessment of educational achievements with this kind of statistics could be questionable. Finally, it should be noted that, even if we were able to know the average years of schooling within each incomplete level, our estimates of MYS would be still biased. The reason is that this methodology does not take into account the right censoring of census data. All individuals classified as having completed tertiary  education are supposed to have finished university studies, but it is conceivable that some of them may have been enrolled in masters or Ph.D programs. Therefore, assuming that individuals within this group received the same years of schooling may bias downwards MYS estimates.

In this paper, we explore a more nuanced alternative, which considers the continuous nature of the educational variable. This `continuous approach' allows us to impose more plausible assumptions about the distribution of years of schooling within each level of education, and to take into account the right censoring of the data in the estimation. We argue in this paper that a flexible parametric specification must be considered. The reason is that this functional form allows us to model that the distribution of years of schooling in a developed country may be different from the shape exhibited in a developing economy. At the same time, the shape of the distribution might suffer important transformations over time. More specifically, we employ the generalized gamma (GG) distribution to model the time that individuals attend school until they complete the educational cycle or decide to drop out. It should be highlighted from the outset that we use only data on complete levels, thus avoiding any potential bias arising from the estimation of completion rates, which may lead to more reliable estimates of years of schooling.

Years of schooling, however, have been used not only for assessing educational attainment, but also to analyze the evolution of international inequality in education (see Ram, \citeyear*{Ram90}; World Bank, \citeyear*{WDR06}) and to approximate the national distribution of schooling (Thomas et al., \citeyear*{thomas01}; Castello and Domenech, \citeyear*{castello08}; Meschi and Scervini, \citeyear*{meschi}). Nonetheless, since estimates of the global and national distributions of schooling are usually derived from estimates of MYS, it is conceivable that they may also suffer from measurement error issues\footnote{Results of studies dealing with international inequalities in education are expected to represent a lower bound of inequality given that inequality within countries is not being considered. This kind of analysis would provide, \emph{a priori}, valuable information in terms of disparities. Indeed, if an upward trend is observed, we could ensure that the global disparities would also show an ascending pattern. However, since MYS are biased indicators and due to the fact that the bias is not systematic, these results are no longer lower bounds. At the same time, inequality in education at the national level is computed with the same data as those used for MYS on durations and educational attainment rates. As a consequence, these estimates suffer from the same sources of bias as the mean.}. Since education inequality has been often considered as a key factor explaining different aspects of major interest among economists, such as poverty, life expectancy and income inequality (see, e.g. DeGregorio and Lee, \citeyear*{degregorio02}; Castello and Domenech, \citeyear*{castello2008}), we also present in this paper a set of educational inequality measures derived from our MYS series, which should be helpful for a variety of empirical work.

The main contribution of this paper resides, therefore, in the development of a new database of educational attainment and inequality measures of educational outcomes for 142 countries from 1970 to 2010, taking into account the distribution of years of schooling within each level of education\footnote{The full data set including estimates of MYS and inequality measures of educational outcomes is available at www.educationdata.unican.es}. In the following section, we briefly explore previous attempts to measure educational attainment, before describing our methodology to get more reliable estimates of MYS and measures of educational inequality. Thereafter, we introduce the main features of our new data set, and compare our estimates with those provided by Barro and Lee \citeyear{barro2013} and Wail et al. \citeyear{Benaabdelaali12}. We conclude the paper by considering the practical implications of our study.
\newpage
\section{A new approach to estimate educational attainment}

To estimate educational attainment, two types of information are required; firstly, data on educational attainment of each level of education and, secondly, information about their official duration in each country to transform the categorical variable into a cardinal indicator. The first type of data can be drawn from census and surveys compiled by different national and international statistical agencies. Individuals are classified into four different categories: no formal schooling, primary, secondary and tertiary schooling depending on the last year of education attained. When individual data are not available, the proportion of the population in each level of education is extrapolated by making specific assumptions regarding mortality rates and migration trends.

\subsection{Previous measurements of educational attainment}

The development of homogeneous educational attainment estimates for a large number of countries has attracted significant attention from scholars over the last three decades (De la Fuente and Domenech, \citeyear*{delafuente2013}). Several data sets on educational attainment are available now, most of them covering the period after 1960\footnote{With the exception of Morrisson and Murtin (\citeyear*{morrisson09}; \citeyear*{morrisson2013}), who computed estimates of the proportions of the population that attained the four broad educational levels for a very long period that covered the last two centuries. However, this data set includes information for 32 macro-regions rather than at country level.} (Psacharopoulos and Arriagada, \citeyear*{psacharopoulos1986}; Kyriacou, \citeyear*{Kyriacou91}; De la Fuente and Domenech, \citeyear*{delafuente06}; Cohen and Soto  \citeyear*{cohen07}; Lutz et al. \citeyear*{lutz07}; Filmer, \citeyear*{Filmer2010}; Barro and Lee, \citeyear*{barro2013}).  Among these existing data sets, Barro and Lee (henceforth, BL) \citeyear{barro2013} is the most comprehensive database in terms of time and geographical coverage, including information for 146 countries at 5-year intervals over the period 1950-2010.

The last version of BL database improves their previous estimates on MYS, incorporating most of the criticisms pointed out in the literature (see De la Fuente and Domenech, \citeyear*{delafuente06}; Cohen and Soto, \citeyear*{cohen07}). One of the main drawbacks of previous BL series was to consider constant mortality rates across age groups, since aged individuals are expected to have higher mortality rates. The new version of BL series overcomes this potential shortcoming by taking into account the heterogeneity in mortality rates across age groups. An additional improvement is the adjustment of mortality patterns by educational level, which allows to consider the possibility that more educated people would have less probability of dying. The extrapolation method has been also modified to adopt the methodology proposed by Cohen and Soto \citeyear{cohen07}, which seems to have led to more consistent estimates. Particularly, attainment rates for the age cohort $k$ in $t-5$ are calculated from census data of the age cohort $k+5$ in the year $t$, using the mortality patterns of the country.

BL data on educational attainment classify individuals into seven categories: no schooling, incomplete primary, complete primary, incomplete secondary, complete secondary, incomplete tertiary and complete tertiary. At a first stage, educational attainment rates are calculated for the four general categories using census data. Thereafter, these proportions are broken down into complete and incomplete categories using data on completion rates. Completion rates also come from census data, so an extrapolation method is used to estimate missing observations, which may be another potential source of bias.

MYS are then computed using attainment rates and the official durations of educational cycles in each country. UNESCO's Statistical Yearbook reports country-specific duration figures for each educational level For the incomplete levels, however, there is no data for the exact years that a person has been attending a particular educational level, hence an arbitrary value is set for all individuals pertaining to this group. For instance, MYS computed by UNESCO are obtained assuming that people who have not finished a cycle have attained half of its official duration. This assumption can yield under or overestimated figures on MYS depending on the distribution of population within each educational level and, more importantly, it is not feasible to figure out the direction of the bias.

Educational attainment rates of each educational level represent points of the cumulative distribution function (CDF) of the variable length of schooling. For example, primary education attainment rate in the US or the UK would be interpreted as the probability of having less than 6 years of education, which is the current official duration of this educational cycle in both countries. In this sense, we can consider the variable years of schooling as a \emph{discretization} of the continuous variable \textit{time that individuals attend school until leaving the educational system}, which is used to obtain an easily measurable indicator of educational outcomes. Most of the previous studies estimate mean levels of education by linking these points linearly, which implicitly assumes that the population distributes uniformly across academic years. Hence, the mean of that part of the distribution would be given by half of the years of the complete level.

\begin{figure}
\begin{center}
\includegraphics*[scale=0.75]{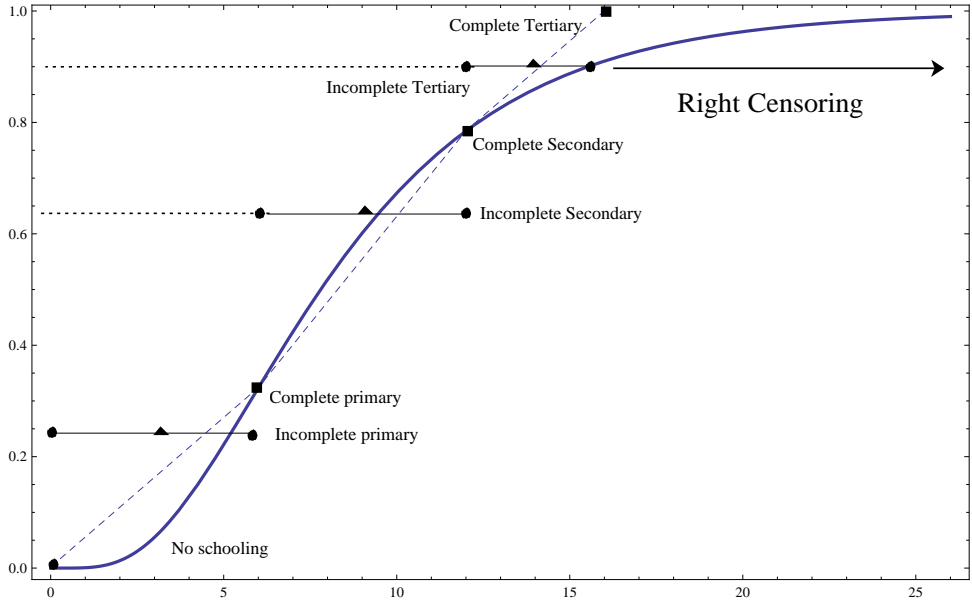}
\caption{\label{FigExample}Sources of bias of mean years of schooling}
\end{center}
\end{figure}

An example may clarify this point; small squares in Figure \ref{FigExample} represent the attainment rates of complete educational levels, for which we know the number of years of schooling associated with the official duration of the cycle. In contrast, this information is not known for incomplete levels. These values are ranged from the maximum years of schooling of the previous educational level to the maximum years of its own cycle, represented in the graph by the black lines at each incomplete level. The dashed line represents the linear interpolation applied in most previous studies. The level of education assigned to the incomplete levels assuming this kind of distribution is given by half of the official duration (triangles placed at the middle of each line). The distribution of time of schooling, which is indeed continuous, would be represented by the solid line. This simple example reveals that it is not possible to find out the direction of the bias. If our choice of the number of years schooling given to the incomplete level lie above (below) the actual mean of the incomplete level, the MYS of the educational cycle would be overestimated (underestimated). The positive or negative bias on the overall MYS is given by the sum of the bias for all levels of education.

It is worth highlighting that even though we were able to know the exact values associated with incomplete attainment rates, our estimates could be still biased, since discrete approaches do not take into account the right censoring of census data. For instance, individuals who have completed a master's degree or higher are considered as having the same number of years of schooling than males or females who have completed just a bachelor's degree or the like, an assumption which may bias downwards MYS estimates.

Following from this, previous attempts to estimate educational inequality based on MYS estimates would suffer from the same sources of bias as the mean (see, for example, two recent studies by Castello, \citeyear*{castello2010}, and Wail et al., \citeyear*{Benaabdelaali12}). In addition, although splitting each educational level in two groups clearly increases the number of available points of the empirical distribution of schooling, we would be still working with grouped data. Hence, computing empirical inequality measures with this kind of limited information may lead to biased results, because the variability within each educational cycle is not considered.

\subsection{Methodology}

In this paper, we use a parametric model to provide a more reliable approximation of the shape of the distribution between the empirical points of the CDF. Let $X$ be a continuous variable representing the time of schooling until either completing the maximum level of education or dropping out school. Let $h^{(j)}_{i}, j=NS, P, S, TI, TC$ be the share of population that attained the educational cycle $j$ (no schooling (NS), primary (P), secondary (S), incomplete tertiary (TI) and complete tertiary (TC)) in country $i$; and denote by $D_i^{(j)}$ its official duration. Educational attainment rates for the first three broad categories represent the frequencies associated to population that attended school during the number of years of official duration or less, $h^{(j)}_{i}=Pr[X\leq D^{(j)}_i]$. As regards the last category, we need to decompose attainment rates of tertiary education to take into account the right censoring of the variable. To do so, we assume that $H_i^{(TI)}$ informs about the proportion of population that attained $D_i^{(T)}$ number of years of schooling or less, and $H_i^{(TC)}$ represents the proportion of population that attained $D_i^{(T)}$ years of schooling or more. Therefore, we only use data on official durations without the need to assign arbitrary values for incomplete levels.

We use BL data on educational attainment (Barro and Lee, \citeyear*{barro2013}) for the four broad educational levels, without considering the decomposition of schooling rates into complete and incomplete, except for the highest category, i.e. tertiary education. Data on the official duration of primary and secondary education are drawn from UNESCO \citeyear{UNESCO}, which includes series from 1970 onwards. We take into account changes in the educational system over time, assuming that those changes apply to the whole population. Following UNESCO's definition,\footnote{For a comprehensive guide about how to use and interpret UNESCO's census data see http://www.uis.unesco.org/Library/Documents/hhsguide04-en.pdf.} illiterate people are considered to have less than 1 year of education. Finally, following previous studies such as Cohen and Soto (\citeyear*{cohen07}), Barro and Lee (\citeyear*{barro2013}) and UNESCO (\citeyear*{UNESCO2013}), we assume that the duration of tertiary education is equal to 4 years in all countries over the whole period\footnote{The choice of four years of schooling is due to the high heterogeneity in the duration of tertiary education programs. On average, duration of completed short-cycle tertiary education (ISCED 5) was 3.7 between 2000 and 2010 (see UNESCO, \citeyear*{UNESCO2013}).}.

After describing the data used to estimate educational attainment, we need to define the modeling strategy. Survival analysis provides a suitable framework to model the time that individuals attend the school until leaving the educational system. This methodology has been extensively used to model the probability of dropping out school (see, for example, Arulampalam et al., \citeyear*{arulampalam04}; Plank et al., \citeyear*{plank05}). This is the first study, to our knowledge, applying survival techniques to model the overall distribution of education.

The main assumption of our methodology is that the same functional form is used to model the distribution of years of schooling in all countries over the whole period. Among the whole range of alternatives\footnote{For a comprehensive review on this topic, we refer the reader to Marshall and Olkin (\citeyear*{Marshall07}).}, the generalized gamma (GG) distribution seems to be the most appealing option. The reason is that this parametric model contains most of the distributions commonly used in survival analysis, including the Weibull, the exponential and the gamma distributions. Hence, the GG distribution would converge to any of its special cases if needed. It should be noted that, even when we assume the same model in all cases, the parameter estimates vary across countries and years, thus allowing us to model particular features of the distribution of schooling. One mode is expected in developed countries with compulsory years of schooling, while zero mode distributions are characteristic of developing countries, which present high illiteracy rates.

The GG distribution can be defined in terms of the probability density function (PDF) as follows (Stacy, \citeyear*{Stacy1962}):
\begin{equation}\label{GGcox}
f_{GG}(x;a,\beta,p)=\frac{ax^{ap-1}e^{-(x/\beta)^a}}{\beta^{ap}\Gamma(a)},\;\;x\ge 0,
\end{equation}%
where $\Gamma(a)=\int_0^\infty x^{a-1}e^{-x}dx $ is the gamma function, $\beta>0$ is a scale parameter,
$a>0$ and $p>0$ are shape parameters.

The CDF is given by the following expression:
$$F(x;a,\beta,p)= IG(x^a;p,\beta^a)=\frac{1}{\Gamma(p)}\int_0^{(x/\beta)^a}t^{p-1}e^{-t}dt$$%
where $IG (.)$ denotes the incomplete gamma function.

The distribution of education of each country is estimated by non-linear least squares (NLS). We use information on the share of the illiterate population and attainment rates of primary, secondary, and incomplete and complete tertiary education. The strategy is to minimize the sum of squared deviations between the empirical attainment rates $(h^{(j)})$ and the associated theoretical probabilities for each level of education $(Pr[X\leq D^{(j)}])$, modeled by the CDF of the GG distribution. The right censoring of the educational variable is taken into account in the estimation by modeling the last category using the survival function $(1-F(x))$ instead of the CDF used for the first four levels.

For each country ($i=1,...,142$) and year ($t=1970,..., 2010$), the function to be minimized is given by,

\begin{equation}\label{minRSS}
\min_{a,\beta,p}\sum_{j=1}^{J-1}\left(IG(v_{it}^{(j)};p,\beta^a)-h_{it}^{(j)}\right)^2+ \left(1-IG(v_{it}^{(TC)};p,\beta^a)-h_{it}^{(TC)}\right)^2,
\end{equation}%

where $v_{it}=\left(D_{it}^{(j)}\right)^a$.

Nonlinear regression techniques involve the definition of starting values for the optimization algorithm. The estimation of the GG distribution is characterized by multiple local minima. This fact makes it difficult to ensure that the estimated values are those which globally minimize the residual sum of squares (RSS). To assess the robustness of our estimates, we consider a set of alternative initial values, using an iterative method inspired in the algorithm proposed by Gomes et al. \citeyear{Gomes08}. Let us consider the following relationship between the gamma and the GG distributions,

$$X \sim GG(a,\beta,p),$$   then   $$X^a\sim G(p,\beta^a).$$

We define a grid for parameter $a$ from 0.2 to 20 by steps of 0.2, and compute the moment estimates for the parameters of the gamma distribution of the variable $X^a$. These estimates, along with the grid value, are used as starting values for the optimization algorithm\footnote{We use the package $optim$ in $R$ to find the minimum of Eq.(\ref{minRSS}). BFGS algorithm is implemented by default and L-BFGS is used when this method reports error. The computation of the gradient is done numerically.}. This process is repeated for all values in the grid, taking the estimation that reports the lowest RSS.

Once the parameters are estimated, MYS are obtained by substituting them in the expression of the mean of the GG distribution:
\begin{equation}\label{mean}
\mu=\frac{\beta\Gamma(p+\frac{1}{a})}{\Gamma(p)}.
\end{equation}

In addition to MYS estimates, we present in this new database several education inequality measures which satisfy the Lorenz ordering. The Lorenz curve is a powerful tool to compare and order distributions according to their inequality levels. If these curves do not cross, the closest distribution to the egalitarian line would be declared as less unequal by any inequality measure consistent with the Lorenz ordering.

Following Sarabia and Jorda (\citeyear*{Sarabia2014}), the Lorenz curve can be generally expressed as,
\begin{equation}\label{LCgen}
L(u)=F_{X_{(1)}}(F^{-1}_{X}(u)),
\end{equation}%
where $F^{-1}_{Y}(u)$ denotes the quantile function and $F_{Y_{(1)}}(x)=\int_0^x tf(t)dt$ is the distribution of the first incomplete moment, defined as follows for the GG distribution:
$$F_{X_{(1)}}(x;a,\beta,p)= IG(x^a;p+1/a,\beta^a)$$

Substituting the explicit expressions of these distributions for the GG family in Eq.(\ref{LCgen}), we obtain the Lorenz curve,
\begin{equation*}
L(u;a,p,\beta) =IG\left[\left(IG^{-1}(u;p,\beta^a)\right)^a;p+\frac{1}{a},\beta^a\right],
\end{equation*}%
where $u\in [0,1]$ and $IG^{-1}(.)$ stands for the inverse of the incomplete gamma function.

However, the Lorenz ordering is partial in the sense that not all distributions can be ranked. In these cases, we need to use inequality measures that provide a complete ordering of distributions. The Gini index has been the main indicator used to measure income inequality and its popularity has spread to education variables, whose disparities have been mainly measured using this indicator (Thomas et al., \citeyear*{thomas01}). This index is sensitive to the middle of the distribution, and it does not allow us to change the weight given to differences in specific parts of the distribution. Varying the sensitivity of inequality measures to the bottom or the upper tails is particularly relevant when there is no Lorenz dominance. If two Lorenz curves cross, inequality measures can point out different results depending on their sensitivity to different parts of the distribution. For this reason, we compute an alternative set of inequality measures belonging to the GE family, which includes a sensitivity parameter which sets the importance given to the differences at the upper tail.

To examine the evolution of educational inequality, we compute the GE measures for different parameter values. The mean log deviation (MLD) corresponds to the GE index when the parameter is set to 0, thus being more sensitive to the bottom part of the distribution. The case given by the Theil's entropy measure is equally sensitive to all parts of the distribution, being characterized by a parameter value equal to 1. We also compute the GE measure when the sensitivity parameter is set equal to 2, which gives more weight to differences in higher education.

For the GG distribution, GE measures are given by the following expressions (Jenkins, \citeyear*{jenkins09}; Sarabia et al., \citeyear*{Sarabia15}):

\begin{equation*}
GE(2)= -\frac{1}{2}+\frac{\Gamma(p+\frac{2}{a})\Gamma(p)}{2\Gamma^2(p+\frac{1}{a})},
\end{equation*}
\begin{equation}\label{theil0gg}
T_0=-\frac{\psi(p)}{a}+\log\frac{\Gamma(p+\frac{1}{a})}{\Gamma(p)},
\end{equation}
\begin{equation}\label{theil1gg}
T_1=\frac{\psi(p+\frac{1}{a})}{a}-\log\frac{\Gamma(p+\frac{1}{a})}{\Gamma(p)}.
\end{equation}%
where $\psi(z)=\Gamma'(z)/\Gamma(z)$ denotes the digamma function.

This family of GE measures are additively decomposable in two components that can be obtained using information on the mean, population shares and national inequality measures. The consideration of additively decomposable measures makes the database extremely useful, since it allows for the computation of overall inequality for any combination of countries. Such flexibility would not be achieved if we had focused solely on the Gini index, which involves, in addition to the between-country and the within-country components, an overlapping term that is specific for the particular group of countries under consideration.

The distribution of education of a particular group or a region of countries is defined as a mixture of the national distributions weighted by their population shares. Let $X_i, i=1,...,N$ be the length of schooling in the county $i$ which is assumed to follow a GG distribution given by Eq.(\ref{GGcox}). Then, the regional PDF of education can be expressed as,
\begin{equation}
f(x)=\sum^{N}_{i=1}\lambda_if_i(x),
\end{equation}%
where $\lambda_i$ stands for the population weights of the countries. The regional CDF would be the integral of the PDF given by the previous equation,
\begin{equation*}\label{mixGG}
F(x)=\sum^{N}_{i=1}\lambda_iF_i(x)=\sum^{N}_{i=1}\lambda_iIG(x^a;p_i,\beta_i^{a_i}),
\end{equation*}

The mean of this distribution would yield the regional MYS which are given by,
\begin{equation*}\label{globalMean}
\mu=\sum^{N}_{i=1}\lambda_i\mu_i=\sum^{N}_{i=1}\lambda_i\frac{\beta_i\Gamma(p_i+\frac{1}{a_i})}{\Gamma(p_i)}.
\end{equation*}

Then, the mean of a particular group of countries is given by a weighted average of the national means and the population shares.

The regional estimates of the GE measures can be easily derived by taking advantage of the decomposition of this family. The general expression of the GE measure is given by,
\begin{equation}
I(\theta)=\sum_{i=1}^{N}\lambda_{i}^{1-\theta}s_{i}^{\theta}I_{i}^{(\theta)}+\frac{1}{\theta(\theta-1)}\left(\sum_{i=1}^{N}\lambda_{i}\left(\frac{\mu_i}{\mu}\right)^{\theta}-1\right),
\end{equation}
where $\lambda_i$ and $I_{i}^{(\theta)}$ are the population share and the GE measure of the country $i$. $s_i$ stands for the proportion of mean income of the country $i$ in the overall mean of the group: $s_i=\frac{\lambda_i\mu_i}{\mu}=\frac{\lambda_i\mu_i}{\sum_{i=1}^{N}\lambda_i\mu_i}$.

The especial cases given by the Theil and the MLD can be decomposed as follows:
$$
T_W=\sum_{i=1}^{N}s_iT_i;  T_B=\sum_{i=1}^{N}s_ilog\left(\frac{\mu_i}{\mu}\right),
$$
$$
L_W=\sum_{i=1}^{N} \lambda_iL_i;  L_B=\sum_{i=1}^{N}\lambda_i log\left(\frac{\mu}{\mu_i}\right),
$$
where $T_i$ and $L_i$ are, respectively, the Theil (Eq.(\ref{theil1gg})) and the MDL (Eq.(\ref{theil0gg})) indices of the country $i$.

\subsection{Goodness of fit}

As already mentioned, the reliability of our estimates relies on the assumption that schooling follows a GG distribution. Therefore, before moving onto the analysis of the evolution of educational patterns, we should investigate the goodness of fit (GOF) of the model. It should be noted that, because we have used grouped data, we do not have information about the size of the sample, thus conventional tests of GOF cannot be applied, because the distribution of the statistic is unknown\footnote{Bootstrapping techniques cannot be used in a context of limited information because we do not know the size of the sample, which has a strong impact on the bootstrap p-values.}. As a measure of GOF, we have computed the RSS for each country/year distribution\footnote{Parameter estimates and residual sum of squares for all distributions are available upon request.}. This measure informs about the size of the differences between the observed attainment rates and the estimated ones using the GG distribution. The smaller the RSS, the closer the values of the estimated and the observed educational achievements.

\begin{table}
\begin{center}
\caption{\label{tableGOF2}
Goodness of fit: quartiles of residual sum of squares by year}
\vspace{0.2cm}
\footnotesize{
\begin{tabular}{c c c c c c}
\toprule
	&	year	&	1st Cuartile	&	Median	&	3th Cuartile	\\
\midrule
	&	1970	&	1.58E-05	&	1.99E-04	&	1.07E-03		\\
	&	1980	&	4.65E-05	&	6.57E-04	&	2.07E-03		\\
Total	&	1990	&	2.65E-04	&	1.10E-03	&	3.32E-03	\\
	&	2000	&	2.30E-04	&	1.97E-03	&	5.37E-03	\\
	&	2010	&	7.35E-04	&	2.86E-03	&	6.88E-03	\\
\midrule									
	&	1970	&	4.81E-05	&	4.50E-04	&	1.60E-03	\\
	&	1980	&	1.28E-04	&	9.06E-04	&	3.24E-03	\\
Males	&	1990	&	4.79E-04	&	1.75E-03	&	4.37E-03	\\
	&	2000	&	4.63E-04	&	2.64E-03	&	6.70E-03	\\
	&	2010	&	6.94E-04	&	3.25E-03	&	8.53E-03	\\
\midrule								
	&	1970	&	3.77E-06	&	7.26E-05	&	5.06E-04	\\
	&	1980	&	2.57E-05	&	2.45E-04	&	1.56E-03	\\
Females	&	1990	&	1.14E-04	&	7.42E-04	&	2.43E-03	\\
	&	2000	&	2.32E-04	&	1.21E-03	&	4.86E-03	\\
	&	2010	&	5.14E-04	&	2.62E-03	&	5.94E-03	\\
\bottomrule
\end{tabular}}
\end{center}
\end{table}

We focus, therefore, on the performance of national estimates using the RSS for both, the entire population and disaggregated by gender. A summary of this information is shown in Table \ref{tableGOF2}, which includes the three quartiles of the distribution of this statistic. The first quartile comprises the best 25\% of the fits, the median the best 50\% and the third quartile the 75\%. Overall, RSS values are reasonably good. The mean deviations of the estimated attainment rates and the observed ones are at most 4 percent (resulting in a RSS value of 0.008) in the 75\% of the estimations performed. These preliminary results suggest that the novel approach proposed in this paper fits the observed attainment rates accurately and, hence, yields reliable estimates of MYS and educational inequality measures.

\section{Data set overview}

The new data set on educational attainment includes data for the 142 countries that have complete information over the period 1970 to 2010. Our data set includes information for population aged over-25 and over-15 for both, total population and disaggregated by gender. In this section, we provide an overview of this new data set figures, starting with the estimation of MYS, before presenting estimates on the proposed educational inequality measures. Table  \ref{table2} summarizes the main trends related to MYS of the population aged 15 years and over, by country income level and by sex. Our estimates suggest, on average, that individuals across the globe are completing more years of schooling with every passing decade. Although there still seems to be some differences across countries and regions, one could observe substantial increases in educational attainment, particularly in low and middle income-countries and, most noticeable, for females.

In 1970, among individuals aged 15 and older in high-income countries, males received an average of 6.35 years of education and females acquired 5.49 years; by 2010 the average increased to 11.18 years for men and 10.88 years for women. These estimates not only reflect an upward trend in educational attainment, but also that female years of schooling have increased relative to that of males during the last decades, thus leading to gender convergence in terms of educational attainment in high-income countries. Regarding low and middle-income countries, the average number of years of education is much lower than in high income countries, but the rate of increase is much higher (about 171.12\% \emph{versus} 87.51\%). On average, years of education rose from 2.70 years for males and 1.75 years for women in 1970, to 6.70 years and 5.51 years in 2010. These estimates indicate that, although there is still a significant gender gap in this group of countries, females are getting closer to males in terms of educational attainment, which is contributing to close that gender gap in low and middle-income countries too.

\begin{figure}[t]
\begin{center}
\includegraphics*[scale=0.75]{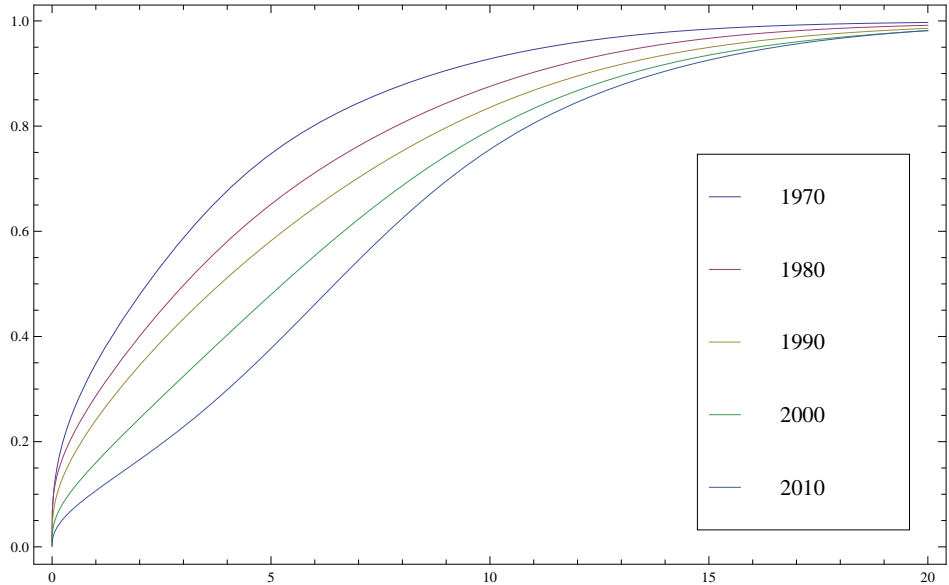}
\caption{\label{Figure6}Global CDF of years of schooling}
\end{center}
\end{figure}

These major advances in educational attainment over time may well reflect those efforts made by countries and international organizations to increase workforce skills and to make progress towards the achievement of universal primary education, one of the Millennium Development Goals (Jorda and Sarabia, \citeyear*{jordawell}). However, it should be noted that, although years of schooling are widely used to measure the educational performance of a country and to make international comparisons, this indicator only informs about the mean of the distribution of educational outcomes. Then, even when countries progress on average, it does not imply that the whole society improves its educational levels. To provide a complete picture of the evolution of global educational achievements, Figure \ref{Figure6} presents the global CDF of educational outcomes for each decade from 1970 to 2010, which is computed as a mixture of the national distributions (Eq.(\ref{mixGG})). This is the first study, to our knowledge, showing the global distribution of schooling instead of limiting the analysis to the frequencies associated to the four levels of education. The most relevant fact is the existence of first order dominance of all years over the preceding ones. This means that the progress in education has been achieved at all levels, a pattern that is rather contrasting with the evolution of income distribution (see Chotickapanich et al., \citeyear*{chotikapanich12}; Jorda et al., \citeyear*{jorda14}).

\begin{table}
\begin{center}

\caption{\label{table2}
Mean years of schooling (population 15 age and over)}
\vspace{0.2cm}
 \scriptsize{
\begin{threeparttable}

\begin{tabular}{l c c c c c c c c c}
\toprule
Year	&	\multicolumn{3}{c}{All countries (142)}			&	\multicolumn{3}{c}{High income (52)}			&	\multicolumn{3}{c}{Low and middle income (90)}\\
\midrule
	&	Total	&	Male	&	Female	&	Total	&	Male	&	Female	&	Total	&	Male	&	Female	\\
\midrule
1970	&	3.4405	&	3.8599	&	3.0399	&	5.8550	&	6.3463	&	5.4927	&	2.2461	&	2.7032	&	1.7526	\\
1975	&	3.8813	&	4.3770	&	3.4125	&	6.5221	&	7.1263	&	6.0885	&	2.6433	&	3.1607	&	2.0853	\\
1980	&	4.3948	&	4.9867	&	3.8614	&	7.1169	&	7.7468	&	6.6863	&	3.2011	&	3.8414	&	2.5542	\\
1985	&	4.8346	&	5.4101	&	4.3004	&	7.7464	&	8.4533	&	7.3231	&	3.6583	&	4.2436	&	3.0153	\\
1990	&	5.2144	&	5.6868	&	4.7292	&	8.4665	&	8.9848	&	8.0342	&	3.9905	&	4.5040	&	3.4254	\\
1995	&	5.7310	&	6.2916	&	5.1944	&	9.1075	&	9.5173	&	8.7763	&	4.5311	&	5.1943	&	3.8660	\\
2000	&	6.2274	&	6.7482	&	5.7644	&	9.7483	&	10.0850	&	9.4740	&	5.0455	&	5.6734	&	4.4682	\\
2005	&	6.7449	&	7.3003	&	6.2449	&	10.3871	&	10.6692	&	10.2262	&	5.5897	&	6.2722	&	4.9339	\\
2010	&	7.2189	&	7.6964	&	6.7904	&	10.9786	&	11.1775	&	10.8810	&	6.0898	&	6.6890	&	5.5168	\\
$\Delta$1970-2010	&	109.82	&	99.39	&	123.38	&	87.51	&	76.13	&	98.10	&	171.12	&	147.45	&	214.78	\\

(\%)	&		&		&		&		&		&		&		&		&		\\

\bottomrule
\end{tabular}

\begin{tablenotes}
\item Note: Data computed for the beginning of each five-year period.
\end{tablenotes}
\end{threeparttable}

}
\end{center}
\end{table}

\begin{figure}
\begin{center}
\includegraphics*[scale=0.75]{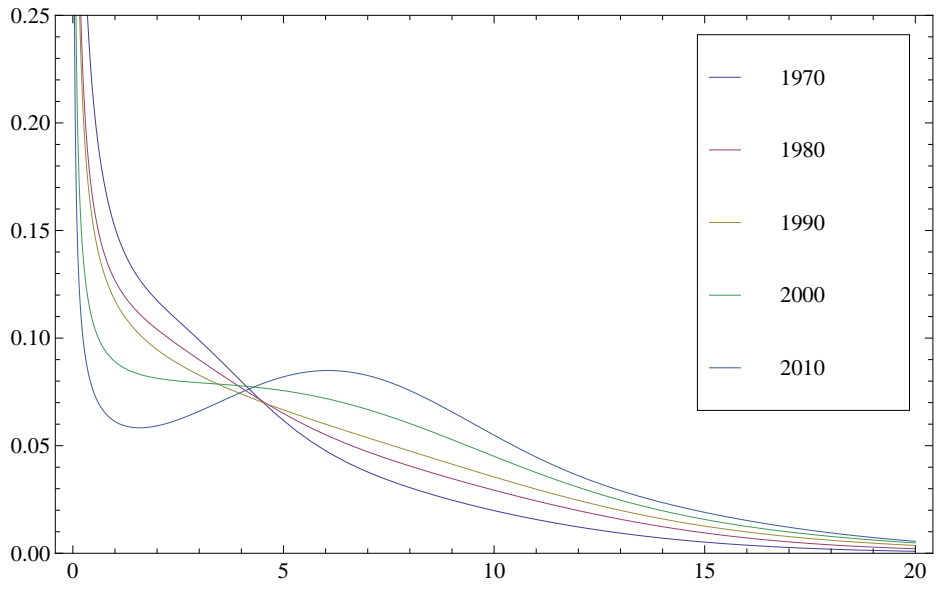}
\caption{\label{Figure2}Global PDF of years of schooling}
\end{center}
\end{figure}

To provide a clearer picture on the evolution of each part of the distribution, Figure \ref{Figure2} shows the global PDF of education. Our results characterize education by a zero mode distribution during almost the whole period. This shape is driven by the high proportion of illiterate population, which is particularly important in developing countries of Saharan-Africa and Asia (United Nations, \citeyear*{MDG14}). Within the last decade, the shape of the distribution changed completely. A new mode emerged around six years of schooling while the proportion of illiterate individuals decreased substantially. In fact, the estimated proportion of people with less than one year of schooling fell from 35 percent to 12 percent over the last 40 years, which involves a reduction of 66 percent. Interestingly, the proportion of people that attained primary education decreased from 40 to 27 percent of the world population. The rationale behind this finding is that the probability mass moved away from the illiterate category to primary schooling and, at the same time, from primary to secondary schooling. Indeed, secondary schooling rates substantially increased over the last decades, due to the convergence process in compulsory years of schooling (Murtin and Viarengo, \citeyear*{murtin11}) and the promotion of education in East Asia (Baker and Holsinger, \citeyear*{Baker1997}). Finally, tertiary education was characterized by an outstanding increase from 4 percent in 1970 to 16 percent in 2010. The improvement of higher education is mainly promoted by developed countries, although developing countries in Asia and Latin America seem to have played also a key role in the last years.

\begin{figure}
\begin{center}
\includegraphics*[scale=0.75]{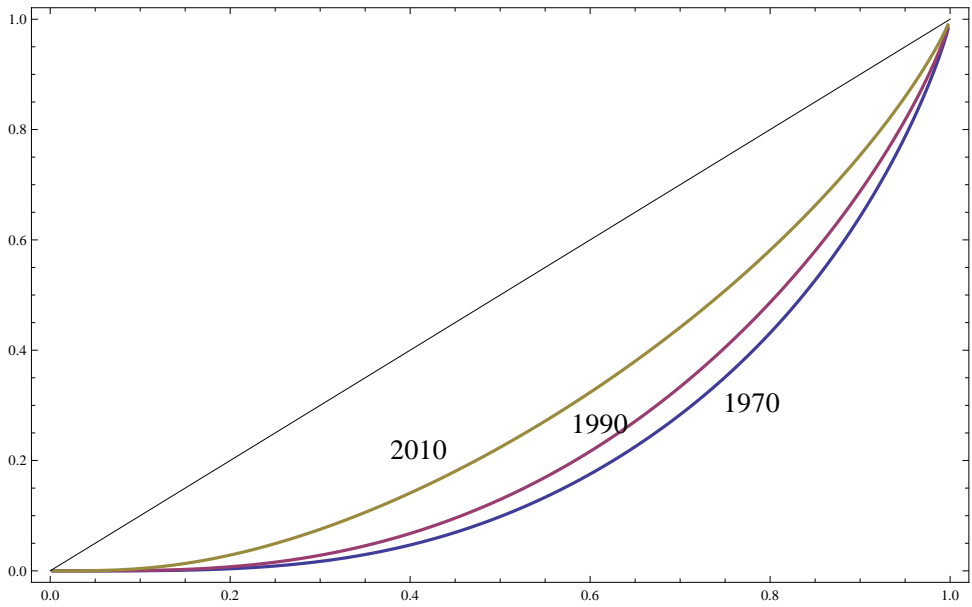}
\caption{\label{FigLC}Global Lorenz curves of education}
\end{center}
\end{figure}

Turning now our attention to the proposed inequality measures, we first investigate the evolution of disparities graphically, before quantifying the global levels of inequality in education. Figure \ref{FigLC} shows the Lorenz curves for years 1970, 1990 and 2010, which reveal a Lorenz dominance pattern over the period under analysis. This result implies that any measure consistent with the Lorenz ordering would reveal a reduction of inequality levels in educational outcomes. Since GE measures satisfy the Lorenz ordering, all these indices concur that inequality levels decreased independently of their sensitivity to the upper tail.

To quantify the reduction in inequality, Table \ref{table3} presents GE measures that (i) are more sensitive to less educated individuals (the MLD - $\theta=0$), (ii) give more importance to higher education ($\theta=2$) or (iii) weight equally all parts of the distribution (the Theil index - $\theta=1$). In line with previous studies, our estimates reveal a substantial decrease in global inequality since the 1970s. This outstanding reduction can be attributable to the decrease of the illiterate population (Morrison and Murtin, \citeyear*{morrisson2013}), although there could be other relevant factors affecting this reduction in educational disparities, such as convergence in compulsory years of schooling (Murtin and Viarengo, \citeyear*{murtin11}) and the expansion of primary education in developing countries, whose attainment rates reached 90 percent in 2012 (United Nations, \citeyear*{MDG14}). Notwithstanding the overall reduction, if we attach more weight to illiterate population, inequality levels tend to increase over the 1970s. The Theil index and the GE(2) measure, in contrast, present a decreasing trend during the whole period, which is consistent with those results reported in Wales et al. \citeyear{Benaabdelaali12}. As pointed out by our estimates, inequality measures might exhibit opposite trends depending on the weighting structure. Then, when there is no Lorenz dominance, the consideration of inequality measures that weights differently each part of the distribution is of main interest.

\begin{table}
\begin{center}
\caption{\label{table3}
Global inequality in education (1970-2010)}
\vspace{0.2cm}
\footnotesize{
\begin{tabular}{c c c c c c c c c c c}
\toprule
Year 	  &	1970	&	1975	&	1980	&	1985	&	1990	&	1995	&	2000	&	2005	&	2010 \\
\midrule
MLD	&	1.8084	&	1.8806	&	1.6931	&	1.3579	&	1.1535	&	0.9921	&	0.8485	&	0.6736	&	0.5691	\\
Between	&	0.1978	&	0.1713	&	0.1414	&	0.1180	&	0.1047	&	0.0910	&	0.0806	&	0.0709	&	0.0624	\\
Within	&	1.6106	&	1.7093	&	1.5517	&	1.2399	&	1.0488	&	0.9011	&	0.7678	&	0.6027	&	0.5067	\\
\midrule
Theil	&	0.5646	&	0.5441	&	0.5030	&	0.4740	&	0.4509	&	0.4000	&	0.3577	&	0.3068	&	0.2756	\\
Between	&	0.1972	&	0.1729	&	0.1418	&	0.1195	&	0.1062	&	0.0903	&	0.0796	&	0.0693	&	0.0614	\\
Within	&	0.3674	&	0.3712	&	0.3612	&	0.3545	&	0.3447	&	0.3098	&	0.2781	&	0.2375	&	0.2142	\\
\midrule
GE (2)	&	0.6227	&	0.5787	&	0.5140	&	0.4804	&	0.4571	&	0.3919	&	0.3457	&	0.2910	&	0.2633	\\
Between	&	0.2288	&	0.1995	&	0.1597	&	0.1334	&	0.1177	&	0.0972	&	0.0847	&	0.0726	&	0.0643	\\
Within	&	0.3939	&	0.3792	&	0.3543	&	0.3470	&	0.3394	&	0.2948	&	0.2611	&	0.2184	&	0.1989	\\

\bottomrule
\end{tabular}}
\end{center}
\end{table}

We exploit the property of decomposability by population subgroups of the GE measures to break down overall inequality into differences in MYS across countries and disparities in educational outcomes within countries. While differences between countries decreased rapidly over the whole period, the fall of within-country inequality is less pronounced. Our estimates also reveal that the differences in education within countries played a predominant role in global inequality, while differences across countries represented a reduced proportion. Moreover, the share of this component in overall inequality steadily decreased over the last 40 years, representing less than 25 percent in 2010. This finding suggests that previous studies on the evolution of international inequality in mean levels of education only analyzed a residual proportion of the global disparities.

\section{Comparison with other data sets}
In this section, we compare our estimates of MYS and educational inequality measures with those series provided by BL and Benaabdelaali et al. (2012) respectively.  Starting with the comparison with BL, Table \ref{table1} presents the average and the standard deviation of these estimates (the continuous approach) along with BL data (the discrete approach) over the period 1970-2010. Although the correlation between both data sets is fairly high, both in levels and in first differences \footnote{Following Krueger and Lindahl (2001), and Cohen and Soto (2007), this correlation between both series may be an indication about the reliability of our estimates. In our case, even 5- and 10-year differences are strongly correlated, which would reinforce the validity of our estimates. It could be argued, however, that the strong level of correlation between both series is driven by the use of the same data on educational attainment, which can correlate positively the measurement errors of both series. While it might foster correlation levels, the use of the same data on educational attainment may not explain itself the high correlation in first differences (Cohen and Soto, \citeyear*{cohen07}) because the method of estimation differs, and even the conceptualization of education is different.}, MYS figures from BL data are considerably higher in all cases.

\begin{table}
\begin{center}
\caption{\label{table1}
Comparison of BL data on average years of schooling (discrete) with the theoretical values of MYS for the GG distribution (continuous). Population over-15.}
\vspace{0.2cm}
\scriptsize{
\begin{threeparttable}
\begin{tabular}{c c c c c c c c}
\toprule
&&\multicolumn{2}{c}{Levels}&  \multicolumn{2}{c}{Correlation}\\
	&		&	Discrete		&	Continuous		&	levels	&	10-year D.	\\
\midrule
Total	&	1970	&	4.3618	(2.6103)	&	3.3018	(1.9772)	&	0.9138	&	-	\\
	&	1980	&	5.4697	(2.7185)	&	4.3103	(2.1893)	&	0.9316	&	0.8888	\\
	&	1990	&	6.4832	(2.6976)	&	5.2677	(2.3026)	&	0.9449	&	0.8153	\\
	&	2000	&	7.4143	(2.7767)	&	6.2596	(2.5757)	&	0.9562	&	0.7813	\\
	&	2010	&	8.3970	(2.7792)	&	7.3504	(2.8311)	&	0.9603	&	0.7372	\\
																	
Males	&	1970	&	4.8816	(2.5729)	&	3.7796	(2.0358)	&	0.9115	&	-	\\
	&	1980	&	5.9923	(2.6170)	&	4.8246	(2.2055)	&	0.9267	&	0.8349	\\
	&	1990	&	6.9564	(2.5407)	&	5.7356	(2.2462)	&	0.9378	&	0.7925	\\
	&	2000	&	7.7970	(2.5951)	&	6.6253	(2.4768)	&	0.9485	&	0.7839	\\
	&	2010	&	8.6425	(2.5983)	&	7.5870	(2.7007)	&	0.9520	&	0.7174	\\
																	
Females	&	1970	&	3.8670	(2.7005)	&	2.8657	(2.0102)	&	0.9282	&	-	\\
	&	1980	&	4.9549	(2.8669)	&	3.8297	(2.2647)	&	0.9368	&	0.9183	\\
	&	1990	&	6.0470	(2.9126)	&	4.8485	(2.4537)	&	0.9496	&	0.8655	\\
	&	2000	&	7.0665	(3.0177)	&	5.9554	(2.7808)	&	0.9624	&	0.7509	\\
	&	2010	&	8.1739	(3.0078)	&	7.1803	(3.0489)	&	0.9652	&	0.7255	\\
\bottomrule
\end{tabular}

\begin{tablenotes}
\item Notes: Standard deviations in parenthesis.
\end{tablenotes}
\end{threeparttable}
}
\end{center}
\end{table}

It should be noted that using overestimated levels of MYS can affect empirical estimates of the role of education on a variety of socio-economics aspects, such as quality of life or human capital formation, particularly if this potential overestimation is not of equal size across all countries. The gap between the discrete and the continuous approach differs across countries, so it is not possible to figure out, \emph{a priori}, how the estimates of the effect of education could be affected\footnote{The computation of reliability ratios is a common practice to check the proportion of the variance of true schooling that is explained by the imperfect indicators of education. The reliability ratio is the coefficient of the bivariate regression of one imperfect measure over another which converges to the variance of true schooling divided by the variance of true schooling plus the variance of the error. However, this result only holds when the errors of the two imperfect measures are uncorrelated. Since our estimates are obtained using BL data on attainment rates, this assumption is extremely implausible.}.

We check now these potential differences through the estimation of a simple production function. It should be highlighted from the outset that our goal is not to analyse those factors affecting income levels, but just to assess whether regression coefficients for MYS differ across both data sets (i.e. the discrete and continuous approaches). To do so, we follow a similar approach to Cohen and Soto \citeyear{cohen07} and estimate the following aggregate production function for country $i$ at time $t$:
\begin{equation}\label{PF}
Y_{it}=A_{it}K_{it}H_{it}
\end{equation}
where $Y_{it}$ is aggregate income or gross domestic product (GDP), $A_{it}$ is the total factor productivity (TFP), $K_{it}$ is physical capital and $H_{it}=h_{it}L_{it}$ ($h_{it}$ denotes human capital per worker; $L_{it}$ represents labour force). Taking logs of (\ref{PF}), diving by $L_{it}$, and assuming that both $log(h_{it})$ and $A_{it}$ are represented, respectively, by the Mincerian approach to human capital\footnote{Although the macro Mincer \citeyear{mincer1974schooling} approach assumes that the logarithm of human capital depends linearly on years of schooling and cumulative work experience, we follow here Krueger and Lindahl \citeyear{krueger2001} and Cohen and Soto \citeyear{cohen07}, and have suppressed the experience term.}, and by the sum of a fixed effect, time effect and the remaining error term, we derive the following equation for the log of GDP per worker in country $i$ at time $t$:
\begin{equation}\label{regmodel}
log(y_{it})= \alpha log(k_{it})+\beta mys_{it}+ \mu_{i}+ \lambda_{t}+ \varepsilon_{it}
\end{equation}
where $mys_{it}$ is the $it$th observation of MYS; $\mu_{i}$ denotes country specific effects; $\lambda_{t}$ represents the specific time effect (common to all countries) and $\varepsilon_{it}$ the remainder disturbance term. In order to compare both data sets, we constructed an unbalanced panel of 132 countries from 1990 to 2010 (five year periods). Information regarding income \emph{per} worker and physical capital \emph{per} worker was obtained from the Penn World Tables, version 8.1 (see Feenstra et al., \citeyear*{pwt}, for a description of the data). We first estimate Eq(\ref{regmodel}) by means of a simple fixed-effects (FE) model. It is conceivable, however, that fixed-effects estimations may result in biased estimates due to the potential endogeneity of physical and human capital. Therefore, to address these potential endogeneity issues, we propose to complement our estimations with a two-stage least squares (2SLS) approach and the generalized method of moments (GMM), using the first two lags of the potential endogenous variables as instruments.

Table \ref{tableESTIMATES} reports FE, 2SLS and GMM estimates for the proposed production function. Starting with the FE specification, the parameter estimate for MYS using our data set is considerably larger than the parameter for MYS using BL data set. These results confirm our expectations about potential differences across estimations using both data sets. In addition, the standard error for MYS using our series is lower than the corresponding model using BL data. We interpret this as a baseline indication that our series may provide more accurate estimates of the effect of MYS on per capita income levels. However, we argued before that the relationship between income levels, physical capital and human capital may be affected for endogeneity issues. Hence, to address this potential source of bias we also report results using 2SLS and GMM estimators. Although the coefficients associated to MYS are slightly smaller when attempting to correct for endogeneity, our previous results hold with both instrumental variable approaches \footnote{Kleibergen-Paap (KP) under-identification and KP weak identification tests suggest that the selected instruments are relevant, and Hansen over-identification tests suggest that those instruments are valid (see Table \ref{tableESTIMATES}).}.

\begin{table}
\begin{center}

\caption{\label{tableESTIMATES}
Comparison of aggregate production function panel estimates between the discrete and continuous approaches}
\vspace{0.2cm}
 \scriptsize{
\begin{threeparttable}

\begin{tabular}{l r r r r r r }
\toprule
	&	\multicolumn{2}{c}{FE}			&	\multicolumn{2}{c}{2SLS}			&	\multicolumn{2}{c}{GMM}			\\

	&	Discrete	&	Continuous	&	Discrete	&	Continuous	&	Discrete	&	Continuous	\\
\midrule
log (k)	&	0.564	&	0.552	&	0.457	&	0.448	&	0.468	&	0.450	\\
	&	(0.043)	&	(0.042)	&	(0.039)	&	(0.039)	&	(0.039)	&	(0.039)	\\
mys	&	0.041	&	0.077	&	0.038	&	0.071	&	0.035	&	0.067	\\
	&	(0.020)	&	(0.015)	&	(0.019)	&	(0.014)	&	(0.019)	&	(0.014)	\\
Observations	&	1070	&	1070	&	808	&	808	&	808	&	808	\\
R-squared	&	0.6317	&	0.6512	&	0.6262	&	0.6387	&	0.627	&	0.6391	\\
K-P LM 	&		&		&	126.21	&	111.174	&	126.21	&	111.174	\\
K-P Wald F	&		&		&	111.425	&	138.546	&	111.425	&	138.546	\\
Hansen J	&		&		&	4.825	&	4.506	&	4.825	&	4.506	\\

\bottomrule
\end{tabular}

\begin{tablenotes}
\item Notes: Dependent and independent variables taken at the beginning of the five-year period. Following Cohen and Soto (2007) mys refers to MYS of population aged 25 years and older. All models include time dummies. Robust standard errors reported in parentheses. Panel 2SLS and GMM models computed with the Stata command \emph{xtivreg2} developed by Schaffer \citeyear{schaffer2012xtivreg2}.
\end{tablenotes}
\end{threeparttable}

}
\end{center}
\end{table}

 Moving now to education inequality measures, Table \ref{TableGini} presents a comparison of our estimates with those provided by Benaabdelaali et al. \citeyear{Benaabdelaali12}, which is the most recent database on educational inequality, covering the period 1950-2010. Their estimates are also computed using BL data on attainment ratios and consider changes in the duration system of the countries. To the best of our knowledge, existing series on educational inequality, such as Benaabdelaali et al. \citeyear{Benaabdelaali12}, only provide Gini coefficient estimates, hence the comparison of our results on inequality focuses solely on this inequality measure \footnote{Morrison and Murtin \citeyear{morrisson2013} also computed Theil indices to analyze global inequality in education. They used, however, data for 32 macro-regions, which prevent us from comparing both sets of estimates on this occasion.}.

 To analyze the shape of the distribution of the educational Gini indices across countries, we present the simple average of the Gini index along with the standard deviation in parenthesis. As expected, the Gini index is higher using the continuous approach, since the discrete approach ignores differences in educational outcomes within levels. Interestingly, the gap between both estimates increases over time; during the 1970s, the `continuous' Gini was only two points higher than the Gini index obtained using the discrete approach, while this gap rose to seven points during the last decade.

\begin{table}
\begin{center}
\caption{\label{TableGini}
Comparison of theoretical and empirical Gini indices of education}
\vspace{0.2cm}
\footnotesize{
\begin{tabular}{c c c c c c c c}
\toprule
&&\multicolumn{2}{c}{Levels}&  \multicolumn{2}{c}{Correlation}\\[-0.2ex]
			& Year& 	Discrete	&	Continuous			&levels	&	10-year D. \\
\midrule
				&	1970	&	0.5283	(0.2286)	&	0.5537	(0.1979)	&	0.9604	&	-	\\
				&	1980	&	0.4700	(0.2114)	&	0.5135	(0.1784)	&	0.9586	&	0.6550	\\
over-15	& 1990	&	0.4132	(0.1887)	&	0.4736	(0.1591)	&	0.9528	&	0.8030	\\
				&	2000	&	0.3568	(0.1779)	&	0.4250	(0.1603)	&	0.9376	&	0.5742	\\
				&	2010	&	0.3089	(0.1585)	&	0.3755	(0.1525)	&	0.8914	&	0.4070	\\
\midrule											
				&	1970	&	0.5698	(0.2420)	&	0.5861	(0.2117)	&	0.9492	&	-	\\
				&	1980	&	0.5174	(0.2285)	&	0.5616	(0.1888)	&	0.9526	&	0.6218	\\
over-25	& 1990	&	0.4581	(0.2079)	&	0.5215	(0.1679)	&	0.9560	&	0.7692	\\
				&	2000	&	0.3923	(0.1966)	&	0.4647	(0.1705)	&	0.9425	&	0.6449	\\
				&	2010	&	0.3382	(0.1811)	&	0.4132	(0.1677)	&	0.8984	&	0.3291	\\
\bottomrule
\end{tabular}}
\end{center}
\end{table}

The correlation coefficient between both data sets is fairly high in levels, being above 90 percent during the whole period. This relation is weaker in 10-year differences, but there is still a strong association between both estimates ranging from 0.8 to 0.6 before 2000. This association weakens substantially during the last decade. We must be cautious about the high correlation observed between both measures because educational Gini indices provided by Wales et al. \citeyear{Benaabdelaali12} are computed using BL data on attainment rates. As a consequence, correlation coefficients may be overestimated because of the positive correlation between the errors of both measures.

\begin{figure}
\begin{center}
\includegraphics*[scale=0.4]{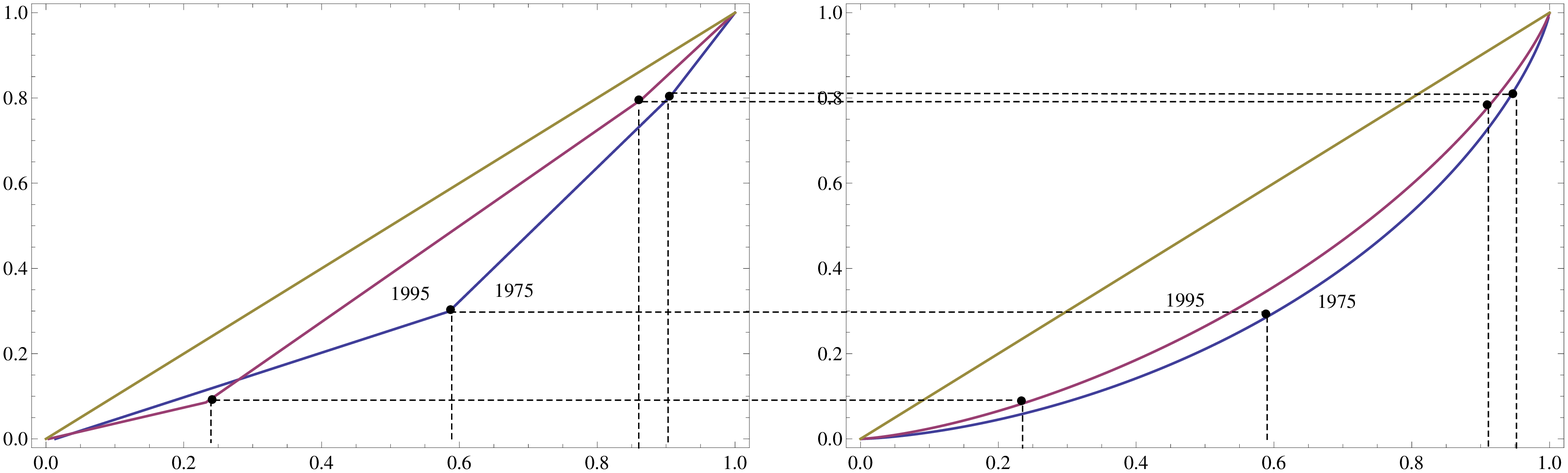}
\caption{\label{FigLatvia}Lorenz curves of education in Latvia using the discrete and the continuous approaches}
\end{center}
\end{figure}

The previous comparison of MYS and inequality measures between the continuous methodology and the discrete approach reveals that the results might differ, depending on the data used. To illustrate this point, Figure \ref{FigLatvia} shows the educational Lorenz curves for Latvia in 1975 and 1995. The Lorenz curve of discrete approach is presented in the left panel. These curves are constructed by plotting on the horizontal axis the accumulated attainment rates and, on the vertical axis, the cumulative proportion of education, which is given by the weighted sum of the attainment rate and the official duration of each level divided by overall MYS. Following Castello and Domenech \citeyear{castello2008}, we do not break down educational levels into complete and incomplete, thus assuming that all individuals classified in some educational level have completed it and hence we assign them the official duration. By applying this method we obtain a step-curve truncated at the horizontal axis due to the assumption that illiterate people have zero years of schooling. On the right side, we plot the curves using the continuous methodology developed in this paper. It should be noted that, by construction, both methodologies produce curves that pass through the black points, as highlighted in the graph by the dashed lines. However, Lorenz dominance is only observed using the continuous approach. These two opposite conclusions are the consequence of assuming different patterns as regards the incomplete levels of education, which lead to different curvatures between the aforementioned black points.

\section {Conclusions}

In this paper, we present a new database for MYS and inequality measures of education, covering 142 countries over the 1970-2010 period, with the aim of providing more accurate estimates than existing series. Previous attempts to measure educational attainment computed MYS as a weighted average of the official duration of each level and attainment rates. Educational cycles were divided into complete and incomplete, whether the educational level has been finished or not. The main shortcoming of these previous studies is the assumption that all individuals who have not completed a particular educational level have the same arbitrary number of years of schooling. This implicit choice may bias the estimates of MYS and, in addition, the direction of the bias cannot be determined \emph{a priori}.

Even when the distribution within each educational stage is unknown, the accuracy of the estimates on MYS and inequality measures rely on a good approximation of the unidentified segments of the empirical distribution. In this study, we deploy a flexible parametric model to estimate the distribution of years of schooling. This methodology allows us to impose more reliable assumptions about the differences in years of schooling within each level of education than assigning the same value to all individuals. In addition, to assess inequality levels in educational outcomes, we focus on the GE family of inequality measures. Different parameter values have been considered, thus changing the sensitivity of these indicators to the upper tail. The use of additively decomposable measures is more convenient than relying solely on Gini coefficients for practical purposes. Indeed, the main advantage of our database is that overall inequality can be computed for any group of countries using only the information included in our data set.

After briefly presenting and discussing main trends on educational attainment and educational inequality over the past decades, we compare our estimates with those based on a discrete approach to measure educational attainment and education inequality. First, we find that average educational outcomes estimated using the discrete approach, such as Barro and Lee \citeyear{barro2013}, are substantially higher than our estimates. In addition,  although the correlation between both data sets is fairly high, we observe that our series may provide more accurate estimates than those relying on the discrete approach. Similar conclusions can be drawn when comparing our educational Gini coefficients and those provided by Wales et al. \citeyear{Benaabdelaali12}, although the correlation between these series in first differences is substantially lower.

In sum, we show that the results of the discrete approach seem to be extremely sensitive to the assumptions made on the number of years of schooling assigned to the incomplete levels. In contrast, the methodology we deploy here avoids relying on such kind of assumptions about the unknown parts of the distribution, which are estimated using a flexible parametric assumption. These improved series on educational attainment and education inequality may be useful, we believe, to improve our understanding of the role of education on different socio-economic aspects, such as quality of life and human capital formation.

\bibliographystyle{econometrica}
\bibliography{Jorda_EducDistr}

\end{document}